\documentclass{iopart}
\usepackage{iopams,amssymb,xspace}
\usepackage{graphicx}

\begin{document}
\newcommand{\Refcite}{Ref.\@\xspace}
\newcommand{\cf}{cf.\@\xspace}
\newcommand{\ie}{i.e.\@\xspace}
\newcommand{\etc}{etc.\@\xspace}
\newcommand{\exc}{\mathrm{exc}}
\newcommand{\n}{\mathrm{n}}

\title[OTBK excess current]{Analytical calculation of the excess current in the OTBK theory}
\author{Gabriel Niebler$^{1,2}$, Gianaurelio Cuniberti$^{2}$ and Tom\'{a}\v{s} Novotn\'{y}$^{1}$}
\address{$^{1}$ Department of Condensed Matter Physics, Faculty of Mathematics and Physics, Charles University, Ke Karlovu 5, 121 16 Prague 2, Czech Republic}
\address{$^{2}$ Institute for Materials Science and Max Bergmann Center of Biomaterials, Dresden University of Technology, D-01062 Dresden, Germany}
\ead{gabriel.niebler@tu-dresden.de}
\date{\today}

\begin{abstract}
We present an analytical derivation of the excess current in
Josephson junctions within the Octavio-Tinkham-Blonder-Klapwijk
theory  for both symmetric and asymmetric barrier strengths. We
confirm the result found numerically by Flensberg \textit{et al.}\
for equal barriers [{\em Physical Review B} {\bf 38}, 8707 (1988)],
including the prediction of negative excess current for low
transparencies, and we generalize it for differing barriers. Our
analytical formulae provide for convenient fitting of experimental
data, also in the less studied, but practically relevant case of the
barrier asymmetry.
\end{abstract}

\pacs{74.45.+c, 74.50.+r, 74.78.Na, 03.75.Lm, 85.25.Cp}

\submitto{\SUST} 

\section{Introduction}

The transport in Josephson junctions
with various different materials
constituting the normal region
has been a very active research field for decades now and continues
to be one. Among the many interesting
transport properties of Josephson junctions are the so-called
subharmonic gap structure (SGS) and the excess current, both of
which were accurately explained by the concept of multiple Andreev
reflections (MAR). The MAR theory was first formulated for
normal--superconducting (NS) interfaces by Blonder, Tinkham, and
Klapwijk (BTK) \cite{Klapwijk1982,Blonder1982} and the BTK theory
and its extensions (especially to the ferromagnetic or non-BCS
superconducting contacts) are still actively used in fitting
experiments \cite{Giubileo2005,Valentine2006,Ren2007} and in
theoretical studies \cite{Xia2002,Linder2008}.

The BTK theory was then extended to full SNS junctions by Octavio,
Tinkham, Blonder and Klapwijk (OTBK) \cite{Octavio1983} and
Flensberg, Bindslev Hansen and Octavio \cite{Flensberg1988}. The
OTKB approach does not keep track of the evolution of the
quasiparticle phase between the interfaces and therefore assumes
complete dephasing in the junction area. This assumption breaks down
for sufficiently small systems such as, e.g., atomic wires and the
fully coherent approach developed in mid 90's \cite{Bratus1995,
Cuevas1996, CuevasPhD} is applicable instead. Nevertheless, the OTBK
theory describes certain systems, such as microbridges, very well
and keeps on being used in the literature both in experimental
\cite{VanHuffelen1993,Kuhlmann1994,Zimmermann1995,Baturina2002,Ishida2005,Bouchiat2009}
as well as theoretical \cite{Pilgram2005} studies. In particular,
its extension to the experimentally relevant situation of asymmetric
junctions was developed and applied in
Refs.~\cite{VanHuffelen1993,Kuhlmann1994,Zimmermann1995}.

In this work we analytically study the excess current in the OTBK
theory. Although the excess current has been derived analytically in
more recent coherent theories \cite{Cuevas1996} it has not been
reported yet in an analytic form in the older incoherent OTBK
approach. We fill in this gap and provide the analytical derivation
of the excess current for incoherent generally asymmetric SNS
junctions described by the OTBK theory. Our formula can be used for
the experimental fitting but it also has implications for the
understanding of the role of coherence within the junction as
discussed in more detail in the concluding section.

\section{OTBK model}
\renewcommand{\arraystretch}{1.7}
\begin{table}[b]
  \begin{center}
    \begin{tabular}{cccr}
      \hline
      $A(E)$                                                & $B(E)$                                           & $T(E)$ &                 \\
      \hline
      $\frac{\Delta^2}{E^2+(1+2Z^2)^2(\Delta^2-E^2)}$       & $\frac{4Z^2(1+Z^2)(\Delta^2-E^2)}
                                                                   {E^2+(1+2Z^2)^2(\Delta^2-E^2)}$             & 0      & for  $\left|E\right|<\Delta$ \\
      \hline
      $\frac{\Delta^2}
             {\left(E+(1+2Z^2)\sqrt{E^2-\Delta^2}\right)^2}$ & $\frac{4Z^2(1+Z^2)(E^2-\Delta^2)}
                                                                {\left(E+(1+2Z^2)\sqrt{E^2-\Delta^2}\right)^2}$ & $\frac
                                                                                                                  {2(E^2-\Delta^2+E(1+2Z^2)\sqrt{E^2-\Delta^2})}
                                                                                                                  {\left(E+(1+2Z^2)\sqrt{E^2-\Delta^2}\right)^2}$  & for $\left|E\right|>\Delta$  \\
      \hline
    \end{tabular}
  \end{center}
  \caption{The reflection and transmission probabilities for an
            \textit{NS}-interface with the dimensionless barrier strength $Z$
            (after \Refcite\cite{Blonder1982}; modified).}
  \label{tab:ABT}
\end{table}
The BTK theory describes the transport through a single
normal--superconducting interface, which is assumed to consist of a
ballistic superconductor in contact with an equally ballistic piece
of normal metal. Scattering can thus only occur at the interface at
$x=0$, which is modelled by a repulsive delta-function potential
$H\delta(x),\,H\geq 0,$ with a dimensionless parameter $Z=H/\hbar
v_F$ ($v_\mathrm{F}$ being the Fermi velocity) that represents the
barrier strength \cite{Blonder1982}. Transport properties are found
by matching the wave functions on either side of this barrier. The
different allowed processes are identified and labelled as follows:
Andreev reflection $A$, normal reflection $B$, and transmission $T$.
The corresponding probabilities $A(E)$, $B(E)$ and $T(E)$ are
expressed as functions of the quasiparticle energy $E$, the
superconducting gap $\Delta$, and the interface's barrier strength
$Z$ (cf.\ \Tref{tab:ABT}). The electrons at the normal side of the
interface are separated into left- and right-moving populations,
represented by the distribution functions $f_\leftarrow(E)$ and
$f_\rightarrow(E)$, respectively. The current through the interface
is then given by
\begin{equation}
  I=\frac{1}{eR_0}\int_{-\infty}^\infty dE \left(f_\rightarrow(E)-f_\leftarrow(E)\right),
  \label{eq:BTK_I}
\end{equation}
where $R_0=\left( 2N(0)e^2 v_\mathrm{F}\mathcal{A} \right)^{-1}$ is
the Sharvin resistance of the perfectly transparent interface
($Z=0$) with $\mathcal{A}$ being the effective cross section of the
contact and $N(0)$ the (single spin) density of states at the Fermi
energy $E_\mathrm{F}$. Blonder, Tinkham and Klapwijk showed
\cite{Blonder1982} that the distribution function for the
left-moving electrons is given by
\begin{equation}
  f_\leftarrow(E)=A(E)\left[1-f_\rightarrow(-E)\right]+B(E)f_\rightarrow(E)+T(E)f_0(E),
  \label{eq:BTK}
\end{equation}
with $f_0(E)$ being the thermal Fermi distribution function
$f_0(E)=1/(1+\exp(\beta(E-\mu)))$, assuming that the incoming
electrons are in thermal equilibrium with their respective leads at
the temperature $1/k_\mathrm{B}\beta$ and the chemical potential $\mu$.

\begin{figure}[bt]
  \begin{center}
    \includegraphics[width=0.8\textwidth]{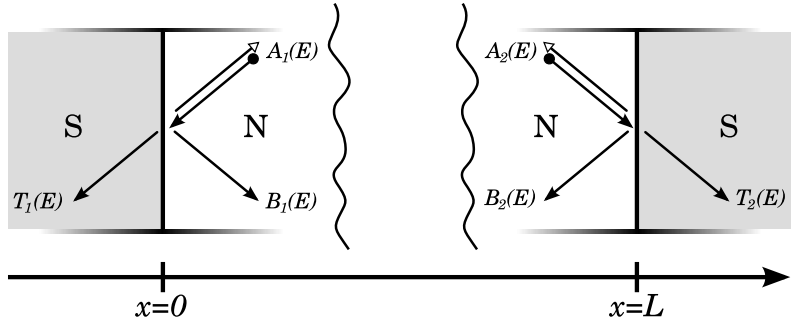}
  \end{center}
  \caption{Schematic representation of an SNS junction, with arrows
           indicating the allowed processes at the interfaces: Andreev
           reflection $A$, normal reflection $B$ and transmission $T$.}
  \label{fig:SNSjunction}
\end{figure}

This description was extended in \Refcite\cite{Octavio1983} to an
SN-interface followed by an NS-interface, \ie to a full Josephson
junction (\cf \Fref{fig:SNSjunction}). We assume the same
superconducting material on both sides, \ie the same superconducting
gap $\Delta$, but different contacts and therefore differing barrier
strenghts. We have interface $1$, located at $x=0$ with barrier
strength $Z_1$, reflection and transmission probablities $A_1(E)$,
$B_1(E)$ and, $T_1(E)$ and interface $2$, at $x=L$ with $Z_2$,
$A_2(E)$, $B_2(E)$ and $T_2(E)$. The distribution functions
$f_\rightleftarrows(E,x)$, which are again to be taken in the normal
region, are also functions of the longitudinal position within the
junction, $x$. Now Eq.~(\ref{eq:BTK}) can be applied to each of the
two interfaces, which yields the following two equations
\cite{Octavio1983}
\begin{equation} \fl
  f_\rightarrow(E,0)=A_1(E)\left[1-f_\leftarrow(-E,0)\right]+B_1(E)f_\leftarrow(E,0)+T_1(E)f_0(E),
  \label{eq:OTBK_fright0}
\end{equation}
\begin{equation} \fl
  f_\leftarrow(E,L)=A_2(E)\left[1-f_\rightarrow(-E,L)\right]+B_2(E)f_\rightarrow(E,L)+T_2(E)f_0(E).
  \label{eq:OTBK_fleftL}
\end{equation}
Note that we only combine distribution functions and not the quantum
states (wavefunctions) at both interfaces. The relative phase of
those states is therefore not considered, which is why the OTBK
model only applies to incoherent junctions.

Since all energies are measured with respect to the local chemical
potential, right-moving quasiparticles with energy $E$ at $x=0$ will
arrive at $x=L$ with energy $E+eV$, while left movers with energy
$E$ at $x=L$ will have energy $E-eV$ at $x=0$. Thus the distribution
functions at the interfaces relate to each other as
\begin{equation}
  f_\rightleftarrows(E,L)=f_\rightleftarrows(E-eV,0).
  \label{eq:OTBK_frightleft}
\end{equation}
Eqs.~(\ref{eq:OTBK_fright0})--(\ref{eq:OTBK_fleftL}) can be combined
to eliminate, e.g., the left-moving part. Using
Eq.~(\ref{eq:OTBK_frightleft}) we can also shift all distribution
functions from $x=L$ to $x=0$ and hence omit the position argument
in the following. The following equation can then be derived
\cite{VanHuffelen1993}
\begin{eqnarray} \fl
  f_\rightarrow(E) &= A_1(E) \Bigl\{ 1 - A_2(-E+eV)\bigl[1-f_\rightarrow( E-2eV)\bigr] \nonumber \\ \fl
                   &\qquad - B_2(-E+eV)f_\rightarrow(-E) - T_2(-E+eV)f_0(-E+eV) \Bigr\} \nonumber \\ \fl
                   &+ B_1(E) \Bigl\{ A_2( E+eV)\bigl[1-f_\rightarrow(-E-2eV)\bigr]   \label{eq:fullfasym} \\ \fl
                   &\qquad + B_2(E+eV)f_\rightarrow(E) + T_2( E+eV)f_0( E+eV) \Bigr\}  \nonumber \\ \fl
                   &+ T_1(E)f_0(E) \nonumber
\end{eqnarray}
that couples $f_\rightarrow(E)$ with $f_\rightarrow(-E)$,
$f_\rightarrow(E-2eV)$ and $f_\rightarrow(-E-2eV)$ and thus gives
rise to an infinite system of linear equations for, say,
$f_\rightarrow(E)$.

\section{Equal barriers}
The simplest case, as far as the barriers are concerned, is the the
case in which both interfaces are characterized by the same barrier
strength $Z_1=Z_2=Z$. For this case an additional relation
$f_\rightarrow(E,0)=1-f_\leftarrow(-E,L)$ was derived in
\Refcite\cite{Flensberg1988} from
Eqs.~(\ref{eq:OTBK_fright0})--(\ref{eq:OTBK_fleftL}) and substituted
into Eq.~(\ref{eq:OTBK_frightleft}), which yields
\begin{equation}
  f_\rightleftarrows(E)=1-f_\leftrightarrows(-E-eV)
  \label{eq:fright2left}
\end{equation}
and greatly simplifies the problem. As before, the suppressed
position arguments imply $x=0$. Using this result we can reformulate
Eq.~(\ref{eq:BTK_I}) to depend on right-movers only
\begin{equation}
  I=\frac{1}{eR_0}\int_{-\infty}^{\infty}dE\left(f_\rightarrow(E)+f_\rightarrow(-E-eV)-1\right).
  \label{eq:OTBK_current}
\end{equation}
Furthermore we can make use of Eq.~(\ref{eq:fright2left}) to
eliminate the distribution functions for left-moving electrons in
Eq.~(\ref{eq:OTBK_fright0}), which yields a significantly simpler
equation than the fully general one from OTBK (\ref{eq:fullfasym}),
namely
\begin{equation} \fl
  f_{\rightarrow}(E) = A(E)f_{\rightarrow}(E-eV)+B(E)\left[1-f_{\rightarrow}(-E-eV)\right]+T(E)f_0(E)
  \label{eq:OTBK_principal}.
\end{equation}
The infinite system of linear equations generated by
Eq.~(\ref{eq:OTBK_principal}) was solved numerically in
\Refcite\cite{Flensberg1988} to obtain subharmonic gap structure and
excess current, but the latter can be obtained analytically
\cite{Niebler2008}, as we reproduce for convenience of the reader in
the following.

\subsection{Normal current}
We shall first calculate the normal current to demonstrate the
course of the derivation and to define some of the quantities used
later on. We introduce the reflection and transmission probabilities
in the normal case
\begin{equation}
  B_\n=\frac{Z^2}{1+Z^2}, \qquad T_\n=1-B_\n=\frac{1}{1+Z^2},
  \label{eq:BnTn}
\end{equation}
which are indeed the limits of $B(E)$ and $T(E)$ for vanishing
$\Delta$, as can be seen from \Tref{tab:ABT}.\footnote{In the normal
state the Andreev reflection coefficient $A(E)$ is identically zero.}
This allows us to rewrite Eq.~(\ref{eq:OTBK_principal}) for
the normal case as
\begin{equation}
  f_\rightarrow^\n(E) = B_\n\left[1-f_\rightarrow^\n(-E-eV)\right]+T_\n f_0(E),
  \label{eq:fn}
\end{equation}
where $f_\rightarrow^\n(E)$ is the right-moving distribution
function for the normal case. We rewrite the above Eq.~(\ref{eq:fn})
for the energy $-E-eV$
\begin{equation}
  f_\rightarrow^\n(-E-eV) = B_\n\left[1-f_\rightarrow^\n(E)\right]+T_\n f_0(-E-eV),
  \label{eq:fn2}
\end{equation}
insert this again in Eq.~(\ref{eq:fn}) and solve for
$f_\rightarrow^\n(E)$, which yields
\begin{eqnarray}
  f_\rightarrow^\n(E) &= \frac{T_\n}{1-B_\n^2}f_0(E)+\frac{B_\n T_\n}{1-B_\n^2}f_0(E+eV) \nonumber\\
         &= \frac{1+Z^2}{1+2Z^2}f_0(E)+\frac{Z^2}{1+2Z^2}f_0(E+eV), \label{eq:fn_final}
\end{eqnarray}
where we have used that $f_0(-E-eV)=1-f_0(E+eV)$. We can write down
the integrand from Eq.~(\ref{eq:OTBK_current}) with these
normal-case distribution functions and simplify it to give
\begin{eqnarray}\fl
  f_\rightarrow^\n(E)+f_\rightarrow^\n(-E-eV)-1 = \frac{1+Z^2}{1+2Z^2}\bigl( f_0(E)+\overbrace{f_0(-E-eV)-1}^{-f_0(E+eV)} \bigr) \nonumber\\
  \qquad + \frac{Z^2}{1+2Z^2}\bigl( f_0(E+eV)+\underbrace{f_0(-E)-1}_{-f_0(-E)} \bigr) \label{eq:OTBK_symintegrand} \\
  = \bigl(f_0(E)-f_0(E+eV)\bigr)/(1+2Z^2), \nonumber
\end{eqnarray}
which is easily integrated and yields the familiar result
$I_\n=V/R_\n$, with the normal state resistance $R_\n=(1+2Z^2)R_0$
of the two-interface ballistic sandwich.\footnote{Note, that due to
the ballistic nature of the junction this resistance is {\em not}
just the sum of the two series resistances of the individual
interfaces.}

\subsection{Excess current}
In the superconducting case we are interested in the excess current
$I_\exc$ defined as $I_\exc=I-I_\n$ in the limit
$eV\rightarrow\infty$, which is what we will assume in the rest of this section.
We define $\Delta f_\rightarrow(E) = f_\rightarrow(E) - f_\rightarrow^\n(E)$,
insert this into Eq.~(\ref{eq:OTBK_current}) and substract the normal part
thus arriving at the formula for the excess current
\begin{equation}
  I_\exc =\frac{1}{eR_0}\int_{-\infty}^{\infty}dE\ \big( \Delta f(E) + \Delta f(-E-eV) \big),
  \label{eq:symI_exc}
\end{equation}
where we have dropped the arrows from the notation, as we are only
dealing with right-movers in this section. To calculate $\Delta
f(E)$ we define, similarly to the above, $\Delta B(E) = B(E)-B_\n$
and $\Delta T(E) = T(E)-T_\n$, substitute all these definitions into
Eq.~(\ref{eq:OTBK_principal}), and solve for $\Delta f(E)$ to obtain
\begin{eqnarray} \fl
  \Delta f(E) =& A(E)\bigl[ f_\n(E-eV) + \Delta f(E-eV) \bigr] \nonumber\\ \fl
                           & + \Delta B(E) \bigl[ 1-f_\n(-E-eV) - \Delta f(-E-eV) \bigr]  \label{eq:Deltaf}\\ \fl
                           & - B_\n\Delta f(-E-eV) + \Delta T(E) f_0(E). \nonumber
\end{eqnarray}
Examining Eq.~(\ref{eq:Deltaf}) and keeping in mind that $A(E)$,
$\Delta B(E)$ and $\Delta T(E)$ tend toward zero for $|E|\gg\Delta$,
we see that there exists only a certain energy range $\epsilon$ of
the order of a few multiples of $\Delta$ where $A(E)$, $\Delta
B(E)$, and $\Delta T(E)$ can be considered non-zero such that $f(E)$
will only differ significantly from $f_\n(E)$ within $\epsilon$
around $E=0$ and $E=-eV$. Since we only consider large bias those
two energy regions are well separated. Therefore we can split
$\Delta f(E)$ into one part which is only nonzero for $|E|<\epsilon$
and vanishes for all other energies and one part with the same
properties for $|E+eV|<\epsilon$. We introduce these parts by
writing
\begin{equation}
  \Delta f(E)=\Delta\tilde{f}(E)+\Delta\tilde{f}^{-eV}(E).
  \label{eq:Deltaf_expand}
\end{equation}
This mathematical procedure is fully in line with the physical
intuition that the only changes of the distribution functions
induced by the superconductivity will occur within the
few-$\Delta$-multiples vicinity of the two Fermi energies
of the leads. We can rewrite Eq.~(\ref{eq:Deltaf}) for $\Delta f(E-eV)$ and
we see that for $|E|<\epsilon$ most terms in the right hand side
simply drop out as they include a vanishing multiplier. So we are
left with
\begin{equation}
  \Delta\tilde{f}^{-eV}(E-eV) = -B_\n\Delta\tilde{f}(-E).
  \label{eq:Deltarel}
\end{equation}
We insert Eq.~(\ref{eq:Deltarel}) into Eq.~(\ref{eq:Deltaf}), still
assuming $|E|<\epsilon$, to obtain
\begin{eqnarray} \fl
  \Delta\tilde{f}(E) = & A(E)\bigl[f_\n(E-eV)-B_\n\Delta\tilde{f}(-E)\bigr] \nonumber\\ \fl
                 & + \Delta B(E) \bigl[ 1-f_\n(-E-eV) + B_\n\Delta\tilde{f}(E) \bigr] \label{eq:Deltaf2}\\ \fl
                 & + B_\n^2\Delta\tilde{f}(E) + \Delta T(E) f_0(E). \nonumber
\end{eqnarray}
It turns out to be convenient to consider
the combination $\Delta\tilde{f}(E)+\Delta\tilde{f}(-E)$ in the following, \ie to
symmetrize the problem. Therefore, we rewrite Eq.~(\ref{eq:Deltaf2})
for $\Delta\tilde{f}(-E)$, sum the result with
Eq.~(\ref{eq:Deltaf2}) and solve for
$\Delta\tilde{f}(E)+\Delta\tilde{f}(-E)$ obtaining
\begin{eqnarray}\fl
  \Bigl[1+B_\n\bigl(A(E)-B(E)\bigr)\Bigr]\Bigl(\Delta\tilde{f}(E) + \Delta\tilde{f}(-E)\Bigr)=\nonumber\\
  \!\!\!\!\!\!\!\!\!\!\!\!\!\!\!\!\!\!\bigl(A(E)-\Delta B(E)\bigr)\bigl[f_\n(E-eV)+f_\n(-E-eV)\bigr] + 2\Delta B(E) + \Delta T(E). \label{eq:Deltaf3}
\end{eqnarray}
Obviously, the above Eq.~(\ref{eq:Deltaf3}) again holds only for the
range $|E|<\epsilon$ in which it was derived. The sum
$f_\n(E-eV)+f_\n(-E-eV)$ that turns up on the right-hand side of
Eq.~(\ref{eq:Deltaf3}) can be calculated using
Eq.~(\ref{eq:fn_final}) and the assumption of large bias, \ie
$eV\rightarrow\infty$, so that we obtain
\begin{eqnarray} \fl
  f_\n(E-eV)+f_\n(-E-eV) &= \frac{1+Z^2}{1+2Z^2}\overbrace{\left(f_0(E-eV)+f_0(-E-eV)\right)}^{\longrightarrow2} \nonumber\\
 & + \frac{Z^2}{1+2Z^2}\underbrace{\left(f_0(E)+f_0(-E)\right)}_{1} = \frac{2+3Z^2}{1+2Z^2}. \label{eq:symfn}
\end{eqnarray}
From the condition of probability conservation
$A(E)+B(E)+T(E)=1$, we see that $\Delta T(E)=-A(E)-\Delta B(E)$. We
substitute this into Eq.~(\ref{eq:Deltaf3}) along with
Eq.~(\ref{eq:symfn}) and are left with
\begin{equation} \fl
  \Delta\tilde{f}(E) + \Delta\tilde{f}(-E) = \frac{\bigl(A(E)-\Delta B(E)\bigr)(1+Z^2)} {\Bigl[1+B_\n\bigl(A(E)-B(E)\bigr)\Bigr](1+2Z^2)}. \label{eq:Deltaf+Deltaf}
\end{equation}

We now take the integrand from Eq.~(\ref{eq:symI_exc}) and expand it
by inserting Eq.~(\ref{eq:Deltaf_expand}) to get
\begin{equation} \fl
  \Delta f(E) + \Delta f(-E-eV)=\Delta\tilde{f}(E)+\Delta\tilde{f}^{-eV}(E)+\Delta\tilde{f}(-E-eV)+\Delta\tilde{f}^{-eV}(-E-eV).\label{eq:Deltaf+full}
\end{equation}
The first and the last terms are nonzero around $E=0$ and we can use
the straightforward modification of Eq.~(\ref{eq:Deltarel}) for the
simplification
$\Delta\tilde{f}(E)+\Delta\tilde{f}^{-eV}(-E-eV)=(1-B_\n)\Delta\tilde{f}(E)$.
Analogously, the two middle terms in Eq.~(\ref{eq:Deltaf+full}) are
nonzero around $E=-eV$ and since they appear under the integral
extending over the entire energy range and have strongly localized
support their energy arguments can be shifted so that they are
localized around $E=0$ as well.\footnote{The strongly localized
support of the involved terms is essential for the possibility of
the variable shift and its lack can lead to seemingly paradoxical
results when done formally, e.g., in Eq.~(\ref{eq:OTBK_current}).}
This eventually leads to the relations
$\Delta\tilde{f}^{-eV}(E)+\Delta\tilde{f}(-E-eV)\Rightarrow\Delta\tilde{f}^{-eV}(E-eV)+\Delta\tilde{f}(-E)=(1-B_\n)\Delta\tilde{f}(-E)$.
Putting all the pieces together leaves us with
\begin{eqnarray} \fl
  I_\exc &= \frac{1}{eR_0}\int_{-\infty}^{\infty}dE\ (1-B_\n)\left(\Delta\tilde{f}(E)+\Delta\tilde{f}(-E)\right) \nonumber\\ \fl
         &= \frac{1}{eR_0(1+2Z^2)}\int_{-\infty}^{\infty}dE\ \frac{A(E)-\Delta B(E)} {\left[1+B_\n\left(A(E)-B(E)\right)\right]},
\end{eqnarray}
where we used Eq.~(\ref{eq:Deltaf+Deltaf}) to produce the final
integrand. The analytical integration must be performed for
$\left|E\right|\leq\Delta$ and $\left|E\right|\geq\Delta$,
separately, because $A(E)$ and $B(E)$ take on different functional
forms in these intervals. It can be evaluated using trigonometric or
hyperbolic substitution for the subgap or overgap energies,
respectively. Thus we find the excess current in the symmetric case
is given by
\begin{eqnarray} \fl
  \frac{eI_\exc R_\n}{\Delta} =& 2(1+2Z^2)\mathrm{tanh}^{-1} \left(2Z\sqrt{ (1+Z^2)/(1+6Z^2+4Z^4) }\right) \nonumber \\ \fl
  & \qquad \times \left( Z\sqrt{(1+Z^2)(1+6Z^2+4Z^4)} \right)^{-1}-\frac{4}{3}. \label{eq:my_Iexc}
\end{eqnarray}
\begin{figure}[bt]
  \begin{center}
    \includegraphics[width=6cm]{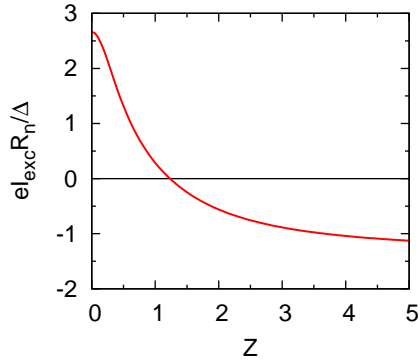}
  \end{center}
  \caption{The excess current of a symmetric, fully incoherent Josephson junction in the OTBK model as a function of the barrier strength $Z$.}
  \label{fig:Iexc_plot}
\end{figure}
The first (and longer) term on the right hand side of
Eq.~(\ref{eq:my_Iexc}) results from the subgap integral, the $-4/3$
term is the contribution of the overgap part. This analytic result
is plotted in \Fref{fig:Iexc_plot} and the comparison with the
earlier numerical result by Flensberg, Bindslev Hansen and Octavio
plotted in Figure~6 of \Refcite\cite{Flensberg1988} shows a nice
agreement.

\section{Differing barrier strengths}
To obtain a similar expression for the excess current in the case of
asymmetric barrier strenghts we need to restart from
Eq.~(\ref{eq:fullfasym}), since Eq.~(\ref{eq:fright2left}) and the
ensuing simplifications, in particular
Eq.~(\ref{eq:OTBK_principal}), cannot be used. The course of the
derivation, however, is very similar to the above. Again we start by
calculating the normal current.
\subsection{Normal current}
Corresponding to Eq.~(\ref{eq:BTK_I}), the normal current is given
by
\begin{equation}
  I_\n=\frac{1}{eR_0}\int_{-\infty}^\infty dE \left(f_\rightarrow^\n(E)-f_\leftarrow^\n(E)\right),
  \label{eq:BTK_normI}
\end{equation}
where $f_\rightarrow^\n(E)$ is defined as above and
$f_\leftarrow^\n(E)$ is its left-moving counterpart. The reflection
and transmission probabilities in the normal case, $B_{\n,i}$ and
$T_{\n,i}$, are defined as above with the additional index
$i\in\{1;2\}$, which indicates the interface in question. From
Eq.~(\ref{eq:fullfasym}) we now find for the right-movers in the
normal case
\begin{equation}
  f_\rightarrow^\n(E) = B_{\n,1} \left[ B_{\n,2}f_\rightarrow^\n(E)+T_{\n,2}f_0(E+eV) \right] + T_{\n,1}f_0(E). \label{eq:fnormright} \\
\end{equation}
Since we also need to consider left-movers in this section, we use
Eqs.~(\ref{eq:OTBK_fright0})--(\ref{eq:OTBK_frightleft}) to derive
the left-moving counterpart to Eq.~(\ref{eq:fullfasym}), which is
not shown for reasons of length, and finally the equivalent of the
above Eq.~(\ref{eq:fnormright}) for left movers, which reads
\begin{equation}
  f_\leftarrow^\n(E) = B_{\n,2} \left[ B_{\n,1}f_\leftarrow^\n(E)+T_{\n,1}f_0(E) \right] + T_{\n,2}f_0(E+eV).  \label{eq:fnormleft}
\end{equation}
We take the integrand from Eq.~(\ref{eq:BTK_normI}) and use
Eqs.~(\ref{eq:fnormright})--(\ref{eq:fnormleft}) to rewrite it as
follows
\begin{eqnarray}
   f_\rightarrow^\n(E)-f_\leftarrow^\n(E) & = \frac{T_{\n,1}T_{\n,2}}{1-B_{\n,1}B_{\n,2}} \left( f_0(E) - f_0(E+eV) \right) \nonumber \\
      & = \frac{1}{1+Z_1^2+Z_2^2} \left( f_0(E) - f_0(E+eV) \right),
\end{eqnarray}
This is easy to integrate and yields $I_\n=V/(\rho_\n R_0)$, with
$\rho_\n=1+Z_1^2+Z_2^2$. Note that $\rho_\n R_0$ simply becomes
$R_\n$ for $Z_1=Z_2=Z$, so we find the normal current from above for
equal barriers again.
\subsection{Excess current}
For the calculation of the excess current we assume large bias once
again and introduce $\Delta
f_\rightarrow(E)=f_\rightarrow(E)-f_\rightarrow^\n(E)$, $\Delta
B_i(E)=B_i(E)-B_{\n,i}$ and $\Delta T_i(E)=T_i(E)-T_{\n,i}$, just
like above in the case of symmetric barriers. Using these relations
we can expand Eq.~(\ref{eq:fullfasym}) and subtract
Eq.~(\ref{eq:fnormright}) to obtain
\begin{eqnarray} \fl
  \Delta f_\rightarrow(E) =& A_1(E) \Bigl\{ 1 - A_2(-E+eV)\bigl[1-f_\rightarrow( E-2eV)\bigr] \nonumber \\ \fl
                           & \qquad - B_2(-E+eV)f_\rightarrow(-E) \nonumber \\ \fl
                           & \qquad - T_2(-E+eV)f_0(-E+eV) \Bigr\}  \nonumber \\ \fl
                           & + B_{\n,1} \Bigl\{ A_2( E+eV)\bigl[1-f_\rightarrow(-E-2eV)\bigr]+B_{\n,2}\Delta f_\rightarrow(E) \label{eq:fulldeltafright} \\ \fl
                           & \qquad + \Delta B_2(E+eV)f_\rightarrow(E) + \Delta T_2(E+eV)f_0(E+eV) \Bigr\}  \nonumber \\ \fl
                           & + \Delta B_1(E) \Bigl\{ A_2( E+eV)\bigl[1-f_\rightarrow(-E-2eV)\bigr] \nonumber \\ \fl
                           & \qquad + B_2(E+eV)f_\rightarrow(E) + T_2( E+eV)f_0( E+eV) \Bigr\}  \nonumber \\ \fl
                           & + \Delta T_1(E)f_0(E). \nonumber
\end{eqnarray}
By the same logic as before we see that $\Delta f_\rightarrow(E)$ is
only nonzero for $|E|<\epsilon$ or $|E+eV|<\epsilon$. Therefore we
split $\Delta f_\rightarrow(E)$ into two parts, just like we did
above and with the same properties
\begin{equation}
  \Delta f_\rightarrow(E)=\Delta\tilde{f}_\rightarrow(E)+\Delta\tilde{f}_\rightarrow^{-eV}(E).
  \label{eq:Deltafright_expand}
\end{equation}
For the remainder of the section we assume small energies
($|E|<\epsilon$), in which case Eq.~(\ref{eq:fulldeltafright}) can
be reduced and solved for $\Delta\tilde{f}_\rightarrow(E)$ to yield
\begin{eqnarray} \fl
  \bigl[1-B_1(E)B_{\n,2}\bigr]\Delta\tilde{f}_\rightarrow(E) &= A_1(E) \bigl[ 1 - B_{\n,2} \bigl( f_\rightarrow^\n(-E)+\Delta\tilde{f}_\rightarrow(-E) \bigr) \nonumber \\
  &\qquad -T_{\n,2}f_0(-E+eV) \bigr] \label{eq:Deltaftilderight} \\
  &+ \Delta B_1(E) \bigl[ B_{\n,2}f_\rightarrow^\n(E)+T_{\n,2}f_0(E+eV) \bigr] \nonumber \\
  &+ \Delta T_1(E)f_0(E) \nonumber
\end{eqnarray}
We rewrite Eq.~(\ref{eq:Deltaftilderight}) for
$\Delta\tilde{f}_\rightarrow(-E)$, sum the result with
Eq.~(\ref{eq:Deltaftilderight}) and solve for
$\Delta\tilde{f}_\rightarrow(E)+\Delta\tilde{f}_\rightarrow(-E)$,
which gives
\begin{eqnarray} \fl
  \bigl[1+B_{\n,2}\bigl(A_1(E)-B_1(E)\bigr)\bigr]\bigl[ \Delta\tilde{f}_\rightarrow(E)+\Delta\tilde{f}_\rightarrow(-E) \bigr] \nonumber \\
  = A_1(E) \bigl[ 2 - B_{\n,2} \bigl( f_\rightarrow^\n(E)+f_\rightarrow^\n(-E) \bigr) \nonumber \\
  \qquad - T_{\n,2}\bigl( f_0(E+eV)+f_0(-E+eV) \bigr) \bigr] \nonumber \\
  + \Delta B_1(E) \bigl[ B_{\n,2}\bigl( f_\rightarrow^\n(E)+f_\rightarrow^\n(-E) \bigr) \label{eq:Deltaftildesum} \\
  \qquad + T_{\n,2}\bigl( f_0(E+eV)+f_0(-E+eV) \bigr) \bigr] \nonumber \\
  + \Delta T_1(E) \underbrace{\bigl( f_0(E)+f_0(-E) \bigr)}_{1} \nonumber
\end{eqnarray}
The sum $f_0(E+eV)+f_0(-E+eV)$ in the above becomes zero for large
bias, which means that the terms explicitly involving $T_{\n,2}$
drop out of Eq.~(\ref{eq:Deltaftildesum}). Furthermore, using
Eq.~(\ref{eq:fnormright}) we can write
\begin{eqnarray} \fl
  f_\rightarrow^\n(E)+f_\rightarrow^\n(-E) = \bigl[ B_{\n,1}T_{\n,2} \overbrace{\bigl(f_0(E+eV)+f_0(-E+eV)\bigr)}^{\rightarrow 0} \nonumber \\
  + T_{\n,1}\underbrace{\bigl(f_0(E)+f_0(-E)\bigr)}_{1}\bigr]/(1-B_{\n,1}B_{\n,2}) = \frac{T_{\n,1}}{1-B_{\n,1}B_{\n,2}},
\end{eqnarray}
further simplifying Eq.~(\ref{eq:Deltaftildesum}), which can now be
written as
\begin{equation} \fl
  \Delta\tilde{f}_\rightarrow(E)+\Delta\tilde{f}_\rightarrow(-E) = \frac{\rho_1}{\rho_\n}\frac{A_1(E)-\Delta B_1(E)}{1+B_{\n,2}\bigl(A_1(E)-B_1(E)\bigr)},  \label{eq:Deltaftildesum01}
\end{equation}
where $\rho_i=1/T_{\n,i}=1+Z_i^2$ is the dimensionless resistance of
the single $i$-th interface in the normal state. In a similar way
and using the same assumptions, \ie $eV\rightarrow\infty$ and
$|E|<\epsilon$, we can show that
\begin{eqnarray} \fl
  \Delta\tilde{f}_\rightarrow^{-eV}(E-eV)+\Delta\tilde{f}_\rightarrow^{-eV}(-E-eV)
  = -\frac{\rho_2}{\rho_\n}\frac{A_2(E)-\Delta B_2(E)}{1+B_{\n,1}\bigl( A_2(E)-B_2(E) \bigr)}B_{\n,1}. \label{eq:Deltaftildesum02}
\end{eqnarray}

We still need to get the left-moving equivalents of
Eqs.~(\ref{eq:Deltaftildesum01}), (\ref{eq:Deltaftildesum02}), so
first we derive the counterpart to Eq.~(\ref{eq:fulldeltafright})
for the left-movers, which is not shown, because the derivation
follows the earlier pattern and does not deliver new insights. Just
like above we can split $\Delta f_\leftarrow(E)$ up into
\begin{equation}
  \Delta
  f_\leftarrow(E)=\Delta\tilde{f}_\leftarrow(E)+\Delta\tilde{f}_\leftarrow^{-eV}(E).
  \label{eq:Deltafleft_expand}
\end{equation}
As for the right-movers and in much the same way we can show that
for $|E|<\epsilon$ and $eV\rightarrow\infty$
\begin{equation} \fl
  \Delta\tilde{f}_\leftarrow(E)+\Delta\tilde{f}_\leftarrow(-E) = \frac{\rho_1}{\rho_\n}\frac{A_1(E)-\Delta B_1(E)}{1+B_{\n,2}\bigl( A_1(E)-B_1(E) \bigr) }B_{\n,2}
  \label{eq:Deltaftildesum03}
\end{equation}
as well as
\begin{equation} \fl
  \Delta\tilde{f}_\leftarrow^{-eV}(E-eV)+\Delta\tilde{f}_\leftarrow^{-eV}(-E-eV) = -\frac{\rho_2}{\rho_\n}\frac{A_2(E)-\Delta B_2(E)}{1+B_{\n,1}\bigl( A_2(E)-B_2(E) \bigr)}.  \label{eq:Deltaftildesum04}
\end{equation}

\begin{figure}[hbt]
  \begin{center}
    \includegraphics{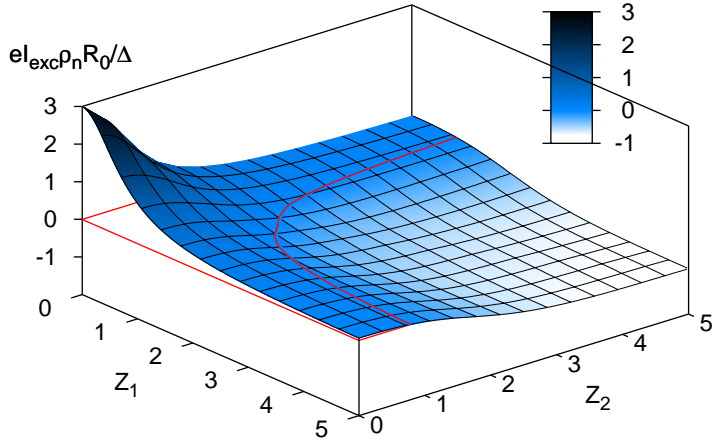}
  \end{center}
  \caption{The excess current of an asymmetric, fully incoherent Josephson junction as a function of the two barrier strengths $Z_1$ and $Z_2$.
  The isoline $I_\exc=0$ is shown in red.}
  \label{fig:Iexc3D}
\end{figure}

The excess current is now given by
\begin{eqnarray} \fl
  I_\exc &= \frac{1}{eR_0}\int_{-\infty}^\infty dE & \bigl( \Delta f_\rightarrow(E)-\Delta f_\leftarrow(E) \bigr) \nonumber \\ \fl
               &= \frac{1}{2eR_0}\int_{-\infty}^\infty dE & \Bigl( \overbrace{\Delta\tilde{f}_\rightarrow(E) + \Delta\tilde{f}_\rightarrow(-E)}^{(\ref{eq:Deltaftildesum01})} + \overbrace{\Delta\tilde{f}_\rightarrow^{-eV}(E) + \Delta\tilde{f}_\rightarrow^{-eV}(-E)}^{(\ref{eq:Deltaftildesum02})'} \label{eq:Iexc_expanded} \\ \fl
               & &- \underbrace{\bigl[\Delta\tilde{f}_\leftarrow(E) + \Delta\tilde{f}_\leftarrow(-E)\bigr]}_{(\ref{eq:Deltaftildesum03})} - \underbrace{\bigl[\Delta\tilde{f}_\leftarrow^{-eV}(E) + \Delta\tilde{f}_\leftarrow^{-eV}(-E)\bigr]}_{(\ref{eq:Deltaftildesum04})'} \Bigr). \nonumber
\end{eqnarray}
The braces and brackets in Eq.~(\ref{eq:Iexc_expanded}) indicate
which terms in the expression correspond to which one of the above
equations. The primed brackets are shifted in energy, which does not
matter to the final result, since the integral extends over the
entire energy range and the integrands have strongly localized
support. Finally we can express the excess current as
\begin{equation} \fl
  I_\exc = \frac{1}{2eR_0}\int_{-\infty}^\infty dE \Biggl\{\frac{\rho_1}{\rho_\n}\frac{A_1(E)-\Delta B_1(E)}{1+B_{\n,2}\bigl( A_1(E)-B_1(E) \bigr)} (1-B_{\n,2})+ \{1\leftrightarrow 2\} \Biggr\}. \label{eq:Iexc_expanded_substituted}
\end{equation}
The integral in Eq.~(\ref{eq:Iexc_expanded_substituted}) can be
solved and the result for $Z_1>Z_2$ is given by
\begin{eqnarray} \fl
  \frac{eI_\exc \rho_\n R_0}{\Delta} &= 2\rho_\n\mathrm{tanh}^{-1}\left(2Z_1\sqrt{\rho_1/\big(2Z_2^2+(1+2Z_1^2)^2\big)}\right) \nonumber \\ \fl
  &\qquad \times \left(Z_1\sqrt{\rho_1\big(2Z_2^2+(1+2Z_1^2)^2\big)}\right)^{-1}\nonumber \\ \fl
  &+\left[ \mathrm{tan}^{-1}\sqrt{(Z_1^2-Z_2^2)/\rho_\n} - \mathrm{tanh}^{-1}\sqrt{(Z_1^2-Z_2^2)/\rho_\n}\right] \label{eq:asymIexc_final} \\ \fl
  &\qquad \times \frac{(1+2Z_1^2)(1+2Z_2^2)}{2\sqrt{\rho_\n}(Z_1^2-Z_2^2)^\frac32}-1.\nonumber
\end{eqnarray}
For $Z_1<Z_2$ the excess current is obtained by exchanging $Z_1$ and
$Z_2$ in Eq.~(\ref{eq:asymIexc_final}) and the result is thus
symmetric with respect to the interchange of the two interfaces.
Again, the first term (the first line) on the right hand side of
Eq.~(\ref{eq:asymIexc_final}) corresponds to the sub-gap integral
and it is easy to see how for $Z_1=Z_2=Z$ it becomes the
corresponding term in Eq.~(\ref{eq:my_Iexc}). The remaining two
terms (the second line), which result from the over-gap integral,
converge towards $-4/3$ for $Z_1\rightarrow Z_2$, as we now show by
the Taylor expansion of
$\mathrm{tan}^{-1}(z)=z-\frac13z^3+\frac15z^5-\frac17z^7+\ldots$ and
$\mathrm{tanh}^{-1}(z)=z+\frac13z^3+\frac15z^5+\frac17z^7+\ldots$
resulting in
$\mathrm{tan}^{-1}(z)-\mathrm{tanh}^{-1}(z)=-\frac23z^3+O(z^7)$ so
that the square bracket in the second line of
Eq.~(\ref{eq:asymIexc_final}) tends to
$-\frac23\bigl((Z_1^2-Z_2^2)/\rho_\n\bigr)^{\frac32}$ for
$Z_1\rightarrow Z_2$. Therefore the last two lines of
Eq.~(\ref{eq:asymIexc_final}) reduce to
$-\frac{(1+2Z_1^2)(1+2Z_2^2)}{3\rho_\n^2} - 1$, which simply becomes
$-4/3$  for $Z_1=Z_2$ and we thus recover Eq.~(\ref{eq:my_Iexc}) in
the symmetric case. The full result (\ref{eq:asymIexc_final}) is
plotted in Figure~\ref{fig:Iexc3D} as a function of the two barrier
strengths $Z_{1,2}$. The negative excess current predicted for large
enough normal-state resistance persists for asymmetric junctions
$Z_1\neq Z_2$ with arbitrarily large asymmetry although its
magnitude decreases (also note the prefactor $\rho_\n$ customarily
multiplying the plotted excess current) and, thus, its experimental
observation may be impeded by the asymmetry of real junctions.

\section{Conclusions and outlook}
In this work we have analytically calculated the excess current
within the OTBK theory describing fully incoherent SNS junctions. We
have confirmed previous numerical findings
\cite{Flensberg1988} of negative excess current for large
enough normal-state resistance in junctions with symmetric barriers.
Furthermore, we extended those calculations also to the case of
asymmetric barriers with qualitatively similar results, \ie occurrence
of negative excess current regardless of the asymmetry. Our
formula (\ref{eq:Iexc_expanded_substituted}) can be used also in the
most general case of different superconducting leads, for an
experiment see, e.g.~Ref.~\cite{Zimmermann1995}, where
$\Delta_1\neq\Delta_2$. The presence of two gap values prohibits
further analytical treatment, however, Eq.
(\ref{eq:Iexc_expanded_substituted}) still holds and the integral
can be easily evaluated numerically.

The numerical findings of Ref.~\cite{Flensberg1988} were challenged
in Ref.~\cite{Cuevas1996} (p. 7372, paragraph below Eq.~(30)) and
the negativity of the excess current was interpreted as possibly
stemming from a lack of convergence of the numerical study, \ie from not
reaching the true $eV\to\infty$ limit. Our study clearly
demonstrates that this objection cannot hold since we explicitly
work in the required limit, thus avoiding any finite-$V$ issues. We,
however, do not question the presence of non-trivial issues in the
experimental determination of the excess current related to the
finite voltage and possible heating effects nicely reviewed and
discussed in Ref.~\cite{CuevasPhD}. Apparently, observations of
negative excess current (so called deficit current) have been
reported in experiments \cite{VanHuffelen1993,Kuhlmann1994}.

Nevertheless, we analytically prove the correctness of the old
numerical results \cite{Flensberg1988} predicting negative excess
current within the OTBK theory. The discrepancy with the results of
Ref.~\cite{Cuevas1996} then must stem from the difference of the two
considered models, more specifically, the role of internal coherence
of the junction. While the OTBK theory only considers matching of
the distribution functions between the two interfaces corresponding
to fully incoherent junctions, the Hamiltonian theory of
Ref.~\cite{Cuevas1996} matches the wavefunctions throughout the
whole junction thus fully retaining the coherence within the
junction. The high-voltage properties of the two models differ
even qualitatively, one predicting negative excess current for small
transparencies, the other one not. Another qualitative
difference between OTBK and the fully coherent theory is in their
dependence on the junction asymmetry: While the fully coherent
results in the limit of strong coherent coupling to the leads
($\Gamma_{1,2}\gg\Delta$, relevant for many experiments, e.g.\
\cite{Henrik2006, Henrik2008,Pertti2009}) only depend on the
asymmetry through the total junction resistance \cite{Cuevas1996}, it is
not so in the OTBK case as we can immediately see from our result
for the excess current (Eq.~(\ref{eq:asymIexc_final})) which is
{\em not} a function of $Z_1^2+Z_2^2$ only.

This finding shows that the coherence within the junction plays a
crucial role for the superconducting transport even at finite
voltage bias and therefore the level of decoherence/dephasing within
a junction should be carefully considered when describing a
particular experiment. This effect, \ie nonzero dephasing within the
junction, may be responsible for the experimentally observed
discrepancies between the experiments \cite{Henrik2006,
Henrik2008,Pertti2009} and theoretical predictions \cite{Cuevas1996}
systematically reported recently in Josephson junctions made of
carbon nanotubes. While those discrepancies are currently
interpreted as the superconducting gap renormalization this picture
does not seem to be fully consistent with the positions of the
subharmonic gap structure features, which appear at the positions
determined by the un-renormalized gap value. The dephasing picture
could capture the relevant physical mechanism instead although this
remains an open issue in the currently booming field of
superconducting transport in carbon-allotropes-based Josephson
junctions.

Apart from the obvious usage of our newly derived analytical
formulae for the OTBK excess current to fit experiments
for relatively large and thus fully incoherent junctions, they can
also be used as a limit benchmark of future partially-coherent
theories, possibly relevant for current nanoscale experiments. 
These experiments as well as future devices built from novel low dimensional
materials with peculiar electronic structures, such as graphene
nanoribbons, could realistically be described by existing dephasing
approaches for atomistic models \cite{Pastawski1991,Buttiker2001}
coupled to non-equilibrium transport. The development of such a
partially-coherent theory and the analytical evaluation of its
excess current interpolating between the two limits is our next step.

\section*{Acknowledgments}
We would like to thank Karsten Flensberg and Peter Samuelsson for
stimulating discussions and for drawing our attention to the
relevant facts and literature. The work of GN is supported by the
grant number 120008 of the GA~UK. The work of TN is a part of the
research plan MSM 0021620834 financed by the Ministry of Education
of the Czech Republic. GC and GN acknowledge support by the European
project CARDEQ under contract IST-021285-2.

\end{document}